\documentclass[aps,prd,nofootinbib]{revtex4}
\pdfoutput=1

\usepackage{amsfonts}
\usepackage{amsmath}
\usepackage{amssymb}
\usepackage{anysize}
\usepackage{bold-extra}
\usepackage{color}
\usepackage{enumerate}
\usepackage{fancyhdr}
\usepackage{graphicx}
\usepackage[linktocpage,colorlinks,urlcolor=blue]{hyperref}
\usepackage{mathrsfs}
\usepackage{multirow}
\usepackage[final]{pdfpages}
\usepackage{rotating}
\usepackage{subfig}
\usepackage{textcomp}
\usepackage{url}
\usepackage{verbatim}
\usepackage{wrapfig}
\usepackage{comment}
\usepackage{epigraph}
\usepackage[utf8]{inputenc}
\usepackage{slashed}
\usepackage{bbm}
\usepackage{cancel}
\usepackage{epsf, array, color}
\usepackage[utf8]{inputenc} 
\def\nn{\nonumber}

\def\lsim{\mbox{\raisebox{-.6ex}{~$\stackrel{<}{\sim}$~}}}

\begin{document}

\title{Statistical approach to Higgs couplings in the standard model effective field theory}

\author{Christopher W. Murphy}
\email{cmurphy@quark.phy.bnl.gov}
\affiliation{Department of Physics, Brookhaven National Laboratory, Upton, New York, 11973, USA}

\begin{abstract}
We perform a parameter fit in the Standard Model Effective Field Theory (SMEFT) with an emphasis on using regularized linear regression to tackle the issue of the large number of parameters in the SMEFT.
In regularized linear regression a positive definite function of the parameters of interest is added to the usual cost function.
A cross-validation is performed to try to determine the optimal value of the regularization parameter to use, but it selects the Standard Model (SM) as the best model to explain the measurements.
Nevertheless as proof of principle of this technique we apply it to fitting Higgs boson signal strengths in SMEFT, including the latest Run-2 results.
Results are presented in terms of the eigensystem of the covariance matrix of the least squares estimators as it has a degree model-independent to it.
We find several results in this initial work: the SMEFT predicts the total width of the Higgs boson to be consistent with the SM prediction; the ATLAS and CMS experiments at the LHC are currently sensitive to non-resonant double Higgs boson production.
Constraints are derived on the viable parameter space for electroweak baryogenesis in the SMEFT, reinforcing the notion that a first order phase transition requires fairly low scale Beyond the SM physics.
Finally, we study which future experimental measurements would give the most improvement on the global constraints on the Higgs sector of the SMEFT.
\end{abstract}

\maketitle

\section{Introduction}
\label{sec:intro}
The Higgs boson discovered at the Large Hadron Collider (LHC) very much resembles the one predicted by the Standard Model (SM)~\cite{Khachatryan:2016vau}.
Unfortunately to date no other particles have been discovered at the LHC~\cite{Patrignani:2016xqp}, indicating there is a mass gap between the SM and whatever may lie beyond it.\footnote{Exceptions to this could be a hidden sector with (sub-)GeV particles, possibly related to dark matter~\cite{Battaglieri:2017aum}, or the alignment without decoupling limit of the two-Higgs doublet model~\cite{Gunion:2002zf}.}
Such a separation of scales lends itself to an effective field theory (EFT) treatment, and the Standard Model Effective Field Theory (SMEFT) is a well developed subject~\cite{Brivio:2017vri, Dedes:2017zog}.

An issue when dealing with the SMEFT is the large number of parameters it contains.
There are 2,499 baryon number preserving real parameters at dimension-6~\cite{Alonso:2013hga}, and this number grows exponentially with the number of dimensions~\cite{Henning:2015alf}.
Following the pioneering analysis of Ref.~\cite{Han:2004az} many parameter fits in the SMEFT have been performed~\cite{Corbett:2012ja, Corbett:2013pja, Dumont:2013wma, Grinstein:2013vsa, Ciuchini:2013pca, Pomarol:2013zra, Ellis:2014dva, Wells:2014pga, Baak:2014ora, Trott:2014dma, Ciuchini:2014dea, Ellis:2014jta, Falkowski:2014tna, Durieux:2014xla, Henning:2014wua, Petrov:2015jea, Berthier:2015oma, Banerjee:2015bla, Corbett:2015ksa, deBlas:2015aea, Wells:2015eba, Falkowski:2015jaa, Berthier:2015gja, Englert:2015hrx, Butter:2016cvz, Cirigliano:2016nyn, Berthier:2016tkq, deBlas:2016ojx, Zhang:2016zsp, Brivio:2017bnu, Khanpour:2017cfq, Khanpour:2017inb, Alioli:2017ces, DiVita:2017eyz, Dawson:2017vgm, Jana:2017hqg}.
These more recently analyses have often focused on constraining the Higgs sector of the SMEFT.

In this work we also perform an SMEFT parameter fit, but with an emphasis on a statistical technique aimed at tackling the issue of the large number of parameters.
In particular, the technique we use is a regularized linear regression, where a positive definite function of the parameters of interest is added to the usual cost function.
This prevents the fit from falling into an overfit solution, and, in principle, allows information to be obtained about any number of parameters.
One application of this in particle physics is unfolding a differential cross section from the detector level to the truth level~\cite{Prosper:2011zz, Behnke:2013pga}
Additionally, it is a commonly used technique in machine learning~\cite{Ng, geron2017hands}, and finds various applications in lattice physics, see e.g.~\cite{Lepage:2001ym, Meyer:2016kwb, Hansen:2017mnd}.

As is typically done a cross-validation is performed to try to determine the optimal value of the regularization parameter to use.
However it selects the SM as the best model to explain the experimental measurements.
Nevertheless we persist in studying the SMEFT, contenting ourselves to performing regularized fits with multiple choices for the regularization parameter, and examining how much regulator dependence various quantities have.

As proof of principle of this technique we apply it to fitting Higgs boson signal strengths, including the latest Run-2 results.
Following Ref.~\cite{Berthier:2015gja} we emphasize presenting results in terms of the eigensystem of the covariance matrix of the least squares estimators as it has a degree model-independent to it.
Despite this being an initial study we obtain several useful physics results.
We show the SMEFT predicts the total width of the Higgs boson, which is not yet directly measured, to be consistent with the SM prediction, and that the ATLAS and CMS experiments at the LHC are currently sensitive to non-resonant double Higgs boson production.
We derive constraints on the viable parameter space for electroweak (EW) baryogenesis in the SMEFT, and reinforce the notion that a first order phase transition requires fairly low scale Beyond the SM (BSM) physics.
We study which future experimental measurements would improve the global constraints on the Higgs sector of the SMEFT the most.
This is quantified using ratios of the global determinant parameter (GDP) of Ref.~\cite{Durieux:2017rsg}, which has a natural interpretation in terms of the eigensystem of the covariance matrix.

We expect this technique to be of use to practitioners of both bottom-up and top-down approaches to EFTs.
In the case of the former, this technique could be applied as described in this work to more sophisticated SMEFT predictions as well datasets that included differential measurements, or measurements from outside the Higgs sector of the SMEFT such as EW precision data (EWPD), triple gauge couplings, or flavor measurements.
For the latter, the regularization matrix provides a convenient way to impose a prior assumptions about possible UV physics.
Additionally, another advantage of this approach is that it makes it easy to determine the blind directions in parameter space for a given data set. 

The rest of the paper is organized as follows.
SMEFT predictions for Higgs boson processes including the Higgs trilinear coupling are given in Sec.~\ref{sec:pred}.
Next the fitting procedures used in this work, and the statistical approaches they employ, are described in Sec.~\ref{sec:stat}.
The experimental results used in these fits are compiled in Appendix~\ref{sec:exp}.
Then the results of our fits to the Higgs signal strength measurements are then presented in Sec.~\ref{sec:res} with additional information given in Appendices~\ref{sec:evec} and~\ref{sec:plots}.
Finally, we summarize our findings in Sec.~\ref{sec:sum}.

\section{Standard Model EFT Predictions}
\label{sec:pred}
The Lagrangian of the SMEFT is given by
\begin{equation}
\label{eq:lag}
\mathcal{L}_{SMEFT} = \mathcal{L}_{SM} + \mathcal{L}^{(5)} + \mathcal{L}^{(6)} + \ldots ,
\end{equation}
where the superscript $n$ in the non-SM terms indicates the mass dimension of the operators contained in that term.

The Yukawa couplings and the dimension-6 Wilson coefficients implicit in Eq.~\eqref{eq:lag} are in general matrices in flavor space. 
Additionally, Higgs boson interactions with fermions inherently have a non-trivial flavor structure, and thus it is important that whatever theoretical framework is used to interpret Higgs measurement also have some non-triviality in its flavor structure. 
With these considerations in mind, the number of parameters can be reduced to a somewhat manageable number of 18 by imposing a $U(2)^5$ symmetry under which the first two generations transform as doublets and the third generation as singlets~\cite{Barbieri:2012uh}.\footnote{For applications of this symmetry in semileptonic $B$ physics see e.g.~\cite{Barbieri:2015yvd, Barbieri:2016las, Bordone:2017anc, Buttazzo:2017ixm}.}
In the basis of Ref.~\cite{Grzadkowski:2010es} with an approximate $U(2)^5$ flavor symmetry these operators are: $Q_H$, $Q_{H\Box}$, $Q_{HD}$, $Q_{HG}$, $Q_{HW}$, $Q_{HB}$, $Q_{HWB}$, $Q_{\substack{uH \\ 33}}$, $Q_{\substack{dH \\ 33}}$, $Q_{\substack{e H \\ 33}}$, $Q_{H\ell}^{(3)}$, $Q_{H\ell}^{(1)}$, $Q_{Hq}^{(3)}$, $Q_{Hq}^{(1)}$, $Q_{He}$, $Q_{Hu}$, $Q_{Hd}$, and $Q_{\ell\ell}$.
Operators without a generation label are $U(2)^5$ symmetric.
See Ref.~\cite{Brivio:2017btx} for additional parameter counting along these lines. 

This is a proof of principle work regarding the usefulness of the statistical methods in constraining SMEFT coefficients. 
As such we made an additional simplification with respect to the SMEFT predictions. 
Specifically, we assume that the production or decay of a Higgs boson involving a pair of $W$ or $Z$ boson does not depend on the type of fermion that produces $W$ or $Z$, or the type of fermion the $W$ or $Z$ decays into. 
Clearly both VBF and the associated $Vh$ production mechanisms involve quarks. 
On the other hand, the best results of Higgs decays to $W$s and $Z$s involve leptonic decays of the vector bosons~\cite{Khachatryan:2016vau}.
Given the aforementioned assumptions, a subset of dimension-6 operators from Eq.~\eqref{eq:lag} that is sufficient for our purposes is
\begin{align}
\label{eq:Lgen}
\Delta\mathcal{L}^{(6)} &= \frac{c_H}{v^2} \partial_{\mu}\left(H^{\dagger} H\right)\partial^{\mu} \left(H^{\dagger} H\right) + \frac{c_T}{v^2} \left| H^{\dagger} \overleftrightarrow{D}_{\mu} H\right|^2  + \frac{c_6}{v^2} \left(H^{\dagger} H\right)^3  \\
&+ \frac{\left(H^{\dagger} H\right)}{v^2} \left[c_b \left(\bar{q}_{L3} d_{R3} H\right) + c_t \left(\bar{q}_{L3} u_{R3} \tilde{H}\right) + c_{\tau} \left(\bar{\ell}_{L3} e_{R3} H\right) + \text{h.c.}\right]  \nn \\
&+ \frac{i c_W}{v^2} \left(H^{\dagger} \sigma^i \overleftrightarrow{D}^{\mu} H\right) \left(D^{\nu}W_{\mu\nu}\right)^i + \frac{i c_B}{v^2} \left(H^{\dagger} \overleftrightarrow{D}^{\mu} H\right) \left(D^{\nu}B_{\mu\nu}\right) \nn \\
&+ \frac{i c_{HW}}{v^2} \left(D^{\mu} H\right)^{\dagger} \sigma^i \left(D^{\nu} H\right) W_{\mu\nu}^i + \frac{i c_{HB}}{v^2} \left(D^{\mu} H\right)^{\dagger}  \left(D^{\nu} H\right) B_{\mu\nu} \nn \\
&+ \frac{c_{\gamma}}{v^2} H^{\dagger} H B_{\mu\nu} B^{\mu\nu} + \frac{c_g}{v^2} H^{\dagger} H G_{\mu\nu}^a G^{a \mu\nu} , \nn
\end{align}
where $v = (\sqrt{2} G_F)^{-1/2} \approx 246$~GeV, and $H^{\dagger} \overleftrightarrow{D}_{\mu} H\equiv H^\dagger D_\mu H-(D_\mu H^\dagger) H$.
Only third generation fermions appear on the second line of~\eqref{eq:Lgen}, consistent with our assumption of a $U(2)^5$ flavor symmetry.
A factor of $v^{-2}$ has been extracted from the Wilson coefficients to make the $c_i$ dimensionless.
We will address the effect of different normalizations and UV assumptions later.
This set of 12 parameters is collected into a vector for later convenience
\begin{equation}
\label{eq:params}
\mathbf{c}^{\top} = \{c_H,\, c_T,\, c_{\gamma},\, c_g,\, c_{HW},\, c_{HB},\,  c_{W} \, c_{B},\, c_t,\, c_b,\, c_{\tau},\, c_6\} ,
\end{equation} 
where $\top$ indicates the transpose.

Numerical results for the Higgs boson decay rates in the SMEFT based on Eq.~\eqref{eq:Lgen} are given in~\cite{Manohar:2006gz, Alonso:2013hga, Contino:2014aaa, Brivio:2017vri}.
The contribution to these decay rates from the Higgs trilinear coupling via electroweak loops is given in Ref.~\cite{Degrassi:2016wml}.
Combining these results we have
\begin{align}
\label{eq:sigstr1}
\frac{\Gamma(h \to \tau \tau)}{\Gamma_{SM}(h \to \tau \tau)} &\simeq 1 - 2 c_H - 196 c_{\tau} , \\
\frac{\Gamma(h \to \mu \mu)}{\Gamma_{SM}(h \to \mu \mu)} &\simeq 1 - 2 c_H ,  \nn \\
\frac{\Gamma(h \to b b)}{\Gamma_{SM}(h \to b b)} &\simeq 1 - 2 c_H - 83 c_b - 0.0085 c_t, \nn \\
\frac{\Gamma(h \to c c)}{\Gamma_{SM}(h \to c c)} &\simeq 1 - 2 c_H - 0.015 c_t , \nn
\end{align}
and
\begin{align}
\label{eq:sigstr2}
\frac{\Gamma(h \to W W^*)}{\Gamma_{SM}(h \to W W^*)} &\simeq 1 - 2.02 c_H + 0.72 c_W + 0.61 c_{HW} - 0.057 c_6 , \\
\frac{\Gamma(h \to Z Z^*)}{\Gamma_{SM}(h \to Z Z^*)} &\simeq 1 - 2.02 c_H - 4 c_T + 0.66 c_W + 0.34 c_B + 0.49 c_{HW} \nn \\
&+ 0.26 c_{HB} - 0.24 c_{\gamma} - 0.064 c_6 , \nn \\
\frac{\Gamma(h \to Z \gamma)}{\Gamma_{SM}(h \to Z \gamma)} &\simeq 1 - 2 c_H + 0.12 c_t - 0.12 c_b - 0.0088 c_{\tau} + 1.38 c_W \nn \\
&+ 151 \left(0.16 c_{HW} - 0.32 c_{HB} + 1.58 c_{\gamma} \right) , \nn \\
\frac{\Gamma(h \to \gamma \gamma)}{\Gamma_{SM}(h \to \gamma \gamma)} &\simeq 1 - 2.01 c_H + 0.54 c_t - 0.29 c_b - 0.69 c_{\tau} + 1.66 c_W \nn \\
&- 863 c_{\gamma} - 0.038 c_6  , \nn \\
\frac{\Gamma(h \to g g)}{\Gamma_{SM}(h \to g g)} &\simeq 1 - 2.02 c_H - 2.13 c_t + 4.17 c_b + 589 c_g - 0.051 c_6. \nn
\end{align}
The width of the Higgs boson in the SMEFT is determined based on Eqs.~\eqref{eq:sigstr1},~\eqref{eq:sigstr2} and the SM branching fractions given in Ref.~\cite{deFlorian:2016spz}. 
We find
\begin{align}
\label{eq:sigstrhh}
\frac{\Gamma_h}{\Gamma_{SM, h}} &\simeq 1 - 2.007 c_H - 0.11 c_T - 1.61 c_{\gamma} + 12.3 c_g + 0.18 c_{HW} - 0.067 c_{HB} \\
&+ 0.18 c_W + 0.009 c_B - 0.187 c_t - 47.4 c_b - 12.3 c_{\tau} - 0.018 c_6 . \nn
\end{align}
We take as numerical expression for Higgs boson production in the SMEFT the following:
\begin{align}
\label{eq:sigstr3}
\frac{\sigma(g g \to h)}{\sigma_{SM}(g g \to h)} &\simeq \frac{\Gamma(h \to g g)}{\Gamma_{SM}(h \to g g)} , \\
\frac{\sigma(p p \to j j h)}{\sigma_{SM}(p p \to j j h)} &\simeq 1 - 2.02 c_H - c_T - 0.06 c_{\gamma} + 0.58 c_{HW} + 0.085 c_{HB} \nn \\
&+ 0.71 c_W + 0.085 c_B - 0.05 c_6, \nn \\
\frac{\sigma(p p \to W h)}{\sigma_{SM}(p p \to W h)} &\simeq 1 - 2.03 c_H + 0.61 c_{HW} + 0.72 c_W - 0.081 c_6 , \nn \\
\frac{\sigma(p p \to Z h)}{\sigma_{SM}(p p \to Z h)} &\simeq 1 - 2.04 c_H - 4 c_T - 0.24 c_{\gamma} + 0.49 c_{HW} + 0.34 c_{HB} \nn \\
&+ 0.66 c_W + 0.34 c_B - 0.095 c_6 , \nn \\
\frac{\sigma(p p \to t \bar{t} h)}{\sigma_{SM}(p p \to t \bar{t} h)} &\simeq 1 - 2.11 c_H - 2.01 c_t - 0.29 c_6 \nn .
\end{align}
The relative fractions of $WW$ and $ZZ$ in the vector boson fusion (VBF) production process are approximations based on Ref.~\cite{Djouadi:2005gi}.
Kinematic differences between production and decays modes, e.g. $\Gamma(h \to W W^*)$ versus $\sigma(p p \to W h)$, or production cross sections at different center-of-mass energies, are not taken into account.
Finally, the prediction for double-Higgs boson production in the SMEFT (at 14 TeV and considering only top quarks in the loop) is~\cite{Azatov:2015oxa}
\begin{equation}
\frac{\sigma(g g \to h h)}{\sigma_{SM}(g g \to h h)} \simeq 1 + 4.25 c_H - 469 c_g + 3.7 c_t - 8.8 c_6 .
\end{equation}

\section{Fitting Procedure}
\label{sec:stat}
In this section we discuss the statistics of the two types of fits we perform.
The first is the method of least squares that is ubiquitous in high energy physics. 
We then discuss a variations of this standard approach that can be used to avoid overfitting, regularizing the least squares fit.  

The 55 experimental measurements from Run-1 and Run-2 used included in these fits are compiled in Appendix~\ref{sec:exp}.
The measurements are all Higgs boson signal strengths.
We do not consider differential or boosted Higgs measurements in this work.
In addition, we do not include EWPD, triple gauge coupling, or flavor results in our fit.
Lastly, no attempt is made to take theoretical errors into account in our fit whether they be from the SM prediction or the SMEFT theory error~\cite{Berthier:2015oma, Berthier:2015gja, Berthier:2016tkq}.

We will perform fits to these measurements with and without regularization, and for various choice of which parameters can be non-zero.
A cross-validation test is performed to determine the optimal value of the regularization parameter to use in the fit.

\subsection{Least Squares Review}
We closely follow the presentation of the PDG~\cite{Patrignani:2016xqp} in what follows. 
The chi-squared function in the case of correlated measurements with covariance matrix $V_{ij}$ is
\begin{equation}
\chi^2\left(\mathbf{c}\right) = \left(\mathbf{y} - \boldsymbol{\mu}\left(\mathbf{c}\right)\right)^{\top} V^{-1} \left(\mathbf{y} - \boldsymbol{\mu}\left(\mathbf{c}\right)\right) ,
\end{equation}
where $\mathbf{y}$ is the vector of measurements, $\boldsymbol{\mu}\left(\mathbf{c}\right)$ is the vector of predictions and $\mathbf{c}$ is the vector of parameters to be estimated.

We consider the case where the predicted values are linear functions of the parameters
\begin{equation}
\mu\left(x_i; \mathbf{c}\right) = \sum_{j = 1}^m h_j\left(x_i\right) c_j .
\end{equation}
In the standard case $h_j\left(x\right)$ are $m$ linearly independent functions. 
In addition, $m$ must be less than the number of measurements, $N$. 
Furthermore, at least $m$ of the $x_i$ must be distinct. 

It will be useful in what follows to define $H_{i, j} = h_j\left(x_i\right)$. 
Consider as an example the leading order SMEFT prediction for the $h \to \gamma \gamma$ decay rate from Eq.~\eqref{eq:sigstr2}. 
From this we see that, for instance, $H_{h \to \gamma \gamma, c_{W}} = 1.66$.

The least squares estimators for the parameters $\mathbf{c}$ are defined through $\boldsymbol{\nabla} \chi^2 = 0$,
\begin{equation}
\mathbf{\hat{c}} = \left(H^{\top} V^{-1} H\right)^{-1} H^{\top} V^{-1}\, \mathbf{y} .
\end{equation}
The inverse of the covariance matrix for the estimators is given by Hessian of chi-squared function, $\tfrac{1}{2}\nabla_i \nabla_j \chi^2$, or equivalently
\begin{equation}
U = \left(H^{\top} V^{-1} H\right)^{-1}. 
\end{equation}
Note that for practical purposes we shift the ones in the SMEFT predictions for the Higgs boson's signal strengths into the measured values.

\subsection{Regularized Linear Regression}
As mentioned in the previous subsection, there are a number of conditions that must be satisfied for the standard least squares approach to be used.
This technique is not useful when the covariance matrix of the estimators is ill-defined, which would be the case if, for example, the $H_i$ are not sufficiently unique. 
These requirements can be bypassed by regularizing the least squares fit.
In a regularized linear regression the cost function is augmented with a positive-definite function of the parameters.
In particular, the regularization makes the inverse of the Hessian of the chi-squared function, the Fisher information, well-defined. 

In this work we use the following expression for the chi-square function as it admits a closed form solution for the least squares estimators
\begin{equation}
\chi^2\left(\mathbf{c}\right) = \left(\mathbf{y} - \boldsymbol{\mu}\left(\mathbf{c}\right)\right)^{\top} V^{-1} \left(\mathbf{y} - \boldsymbol{\mu}\left(\mathbf{c}\right)\right) + \mathbf{c}^{\top} \kappa\, \mathbf{c},
\end{equation}
with $\kappa$ being a positive definite matrix.
We primarily use the simple parameterization $\kappa_{ij} = \kappa \delta_{ij}$, which goes by several names: ridge regression, Tikhonov regularization, and $\ell_2$ penalization.
This choice of $\kappa$ is the frequentist analog of adding the same Gaussian prior to each parameter of interest.
Comments on other choices for $\kappa$ are made later.
Another commonly used regularization term is $\beta \sum_i | c_i |$, $\beta > 0$, which is known as Lasso regression or $\ell_1$ penalization.
This is frequentist analog of adding the same Laplacian prior to each parameter of interest.
One may also choose to use elastic net regularization, a linear combination of ridge and Lasso regression.
We save these methods for future work.

In the case of ridge regression, the least squares estimators are given by
\begin{equation}
\label{eq:creg}
\mathbf{\hat{c}} = \left(H^{\top} V^{-1} H + \kappa \mathbbm{1}\right)^{-1} H^{\top} V^{-1}\, \mathbf{y} ,
\end{equation}
with an obvious generalization for different choices of $\kappa_{ij}$.
There is a similar modification to the covariance matrix
\begin{equation}
\label{eq:Ureg}
U = \left(H^{\top} V^{-1} H + \kappa \mathbbm{1}\right)^{-1}. 
\end{equation}

\subsection{Cross-Validation}
In cross-validation the measurements are randomly split into training and validation groups.\footnote{We ignore correlations between measurements during cross-validation.}
The number of measurements assigned to the training group, $n_t$, is varied between 33 and 39 (60\% to 71\%) of the total $n = 55$ measurements.
The test is performed 300 times for each value of $n_t$ considered, and a chi-squared for the training set, $\chi_t^2$, can be computed.
The best-fit parameters are determined from the training set using regularization linear regression with some value of $\kappa$.
These parameters are then used to compute the chi-squared for just the validation set, $\chi^2_v$, which does not include a regularization term.
The optimal choice of $\kappa$ is given by the value which minimizes $\chi_v^2 / n_v$, where $n_v = n - n_t$ is the number of measurements in the validation group.
See e.g. Ref.s~\cite{Ng, geron2017hands} for more information about cross-validation.

\begin{figure}
  \centering
\subfloat{\includegraphics[width=0.49\textwidth]{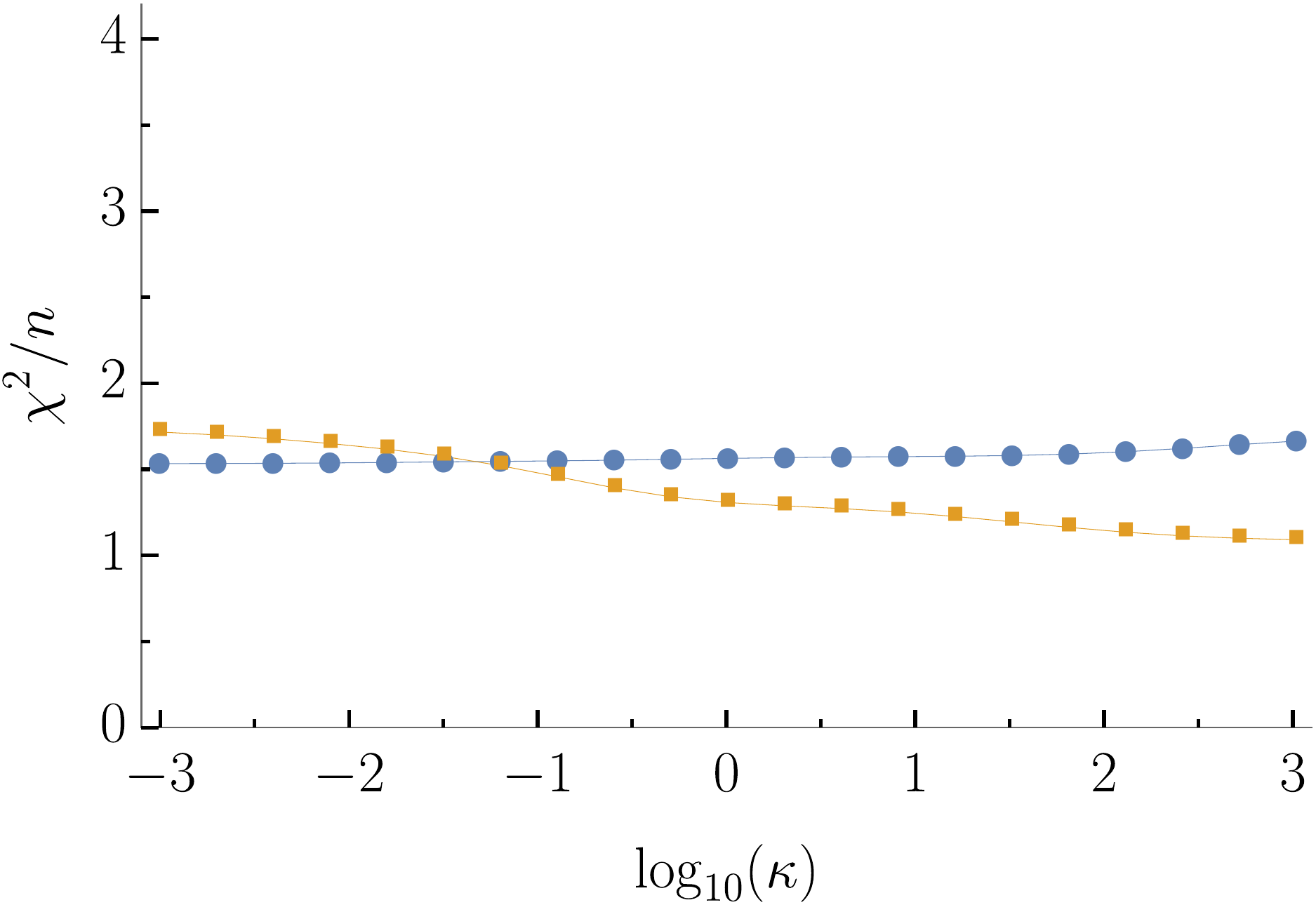}}\,
\subfloat{\includegraphics[width=0.49\textwidth]{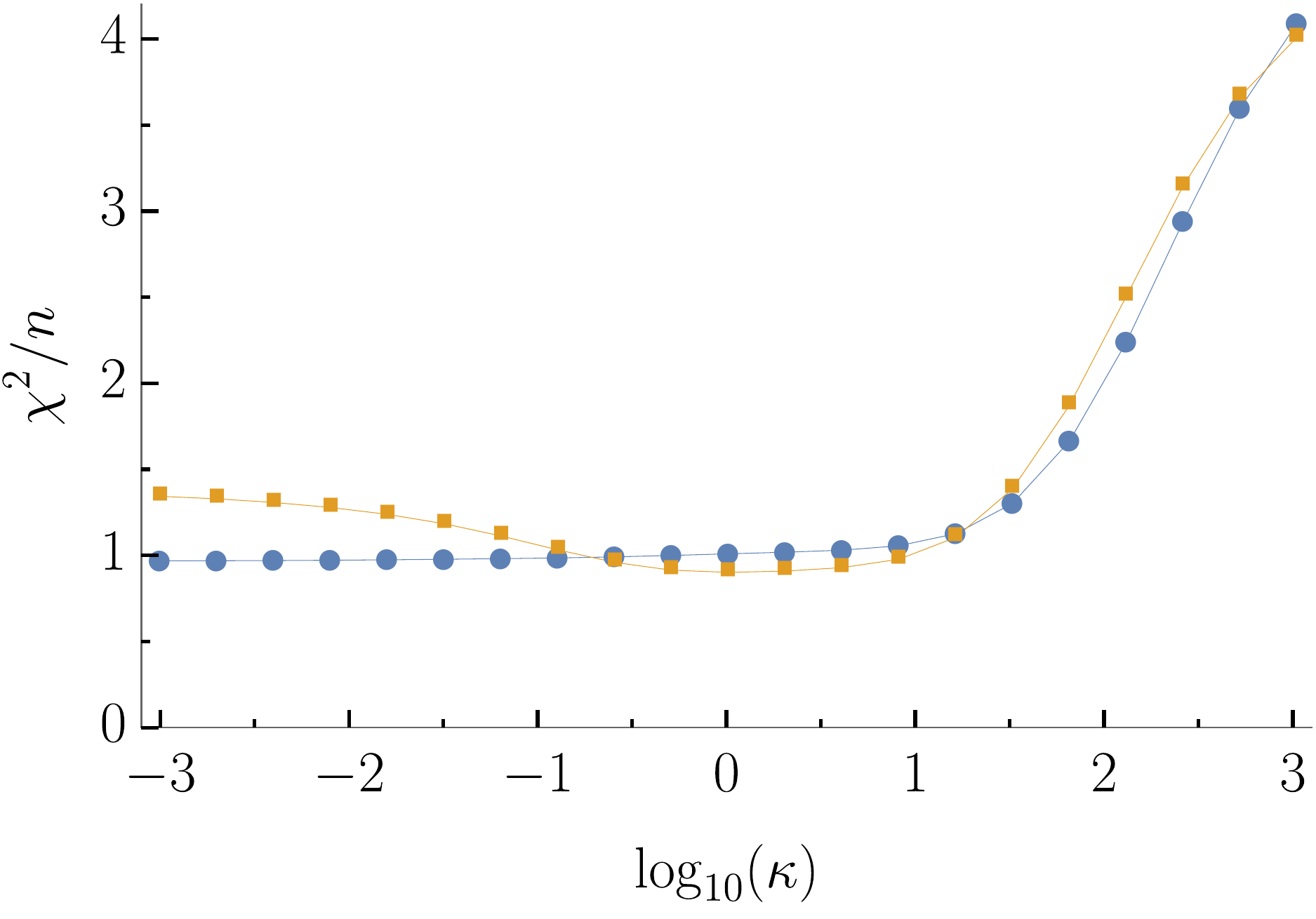}}
 \caption{(Left:) Cross-validation test to determine the optimal value of $\kappa$.
 The average $\chi^2$ per number of measurements is shown as a function of $\kappa$.
 The blue circles and orange squares correspond to the training ($\chi_t^2 / n_t$) and validation ($\chi_v^2 / n_v$)  datasets, respectively.
 The cross-validation selects the SM ($\kappa \to \infty$) as the best model to explain the measurements.
 (Right:) The same as on the left, but with an artificial BSM signal injected.
 Here all $t t h$ signal strength central values have been fixed to 3.0 with their uncertainties left unchanged.
 In this case the cross-validation selects $\kappa \approx 1$ as the best model as it has the lowest (average) value of $\chi_v^2 / n_v$.}
   \label{fig:crossvalidation}
\end{figure}

The average result of the 2100 cross-validation tests for a number of choices of $\kappa$ between $10^{-3}$ and $10^3$ are presented in the left panel of Fig.~\ref{fig:crossvalidation}.
Here the cross-validation selects $\kappa \to \infty$, the Standard Model, as the best model to explain the measurements.
This is an uncommon result as typically the validation curve, the orange curve marked with squares in Fig.~\ref{fig:crossvalidation}, has a local minimum at a finite value of $\kappa$.
Another way to think about this is that the SMEFT may give a lower $\chi^2$ than the SM, but the goodness of fit is still better in the SM.
A similar conclusion, the data prefers the SM over the SMEFT, was drawn in a Bayesian analysis of $b \to s \ell \ell$ observables~\cite{Beaujean:2012uj, Beaujean:2013soa, Meinel:2016grj}

The story would change if there was a (hint of a) signal for BSM physics.
This is illustrated in the right panel of Fig.~\ref{fig:crossvalidation}.
Here we have injected an artificial BSM signal by setting the central values of all the $t t h$ signal strengths to 3.0 while leaving their uncertainties unchanged.
In this case the cross-validation selects $\kappa \approx 1$ as the best model as it has the lowest (average) value of $\chi_v^2 / n_v$.

Interestingly, in both panels of Fig.~\ref{fig:crossvalidation} the cross-validation suggests the data is being underfit.
A hallmark of a model underfitting data is when the validation $\chi_v^2 / n_v$ is comparable to, or smaller than, the training $\chi_t^2 / n_t$.
Both panels of Fig.~\ref{fig:crossvalidation} then suggests that to avoid underfitting a value of $\kappa$ less than one should be chosen.
In any case, as no finite value of $\kappa$ is preferred by the cross-validation, we will typically use two choices for $\kappa$ and see how much regulator dependence there is in our predicted quantities.

\section{Results}
\label{sec:res}

\subsection{Covariance Matrix Eigensystem}
It is instructive to examine the eigensystem of the covariance matrix for the least squares estimators~\cite{Berthier:2015gja}.
There are $k = 1 \ldots 12$ eigenvectors, $W_k = w_{k i} c_i$, normalized such that $|\mathbf{w}_k| = 1$ with $\mathbf{c}$ given in Eq.~\eqref{eq:params}. 
The square root of an eigenvalue, $\sigma_k$, gives the one sigma range on the allowed deviation of eigenvector, $W_k$, from its central value.

\begin{figure}
  \centering
\includegraphics[width=0.65\textwidth]{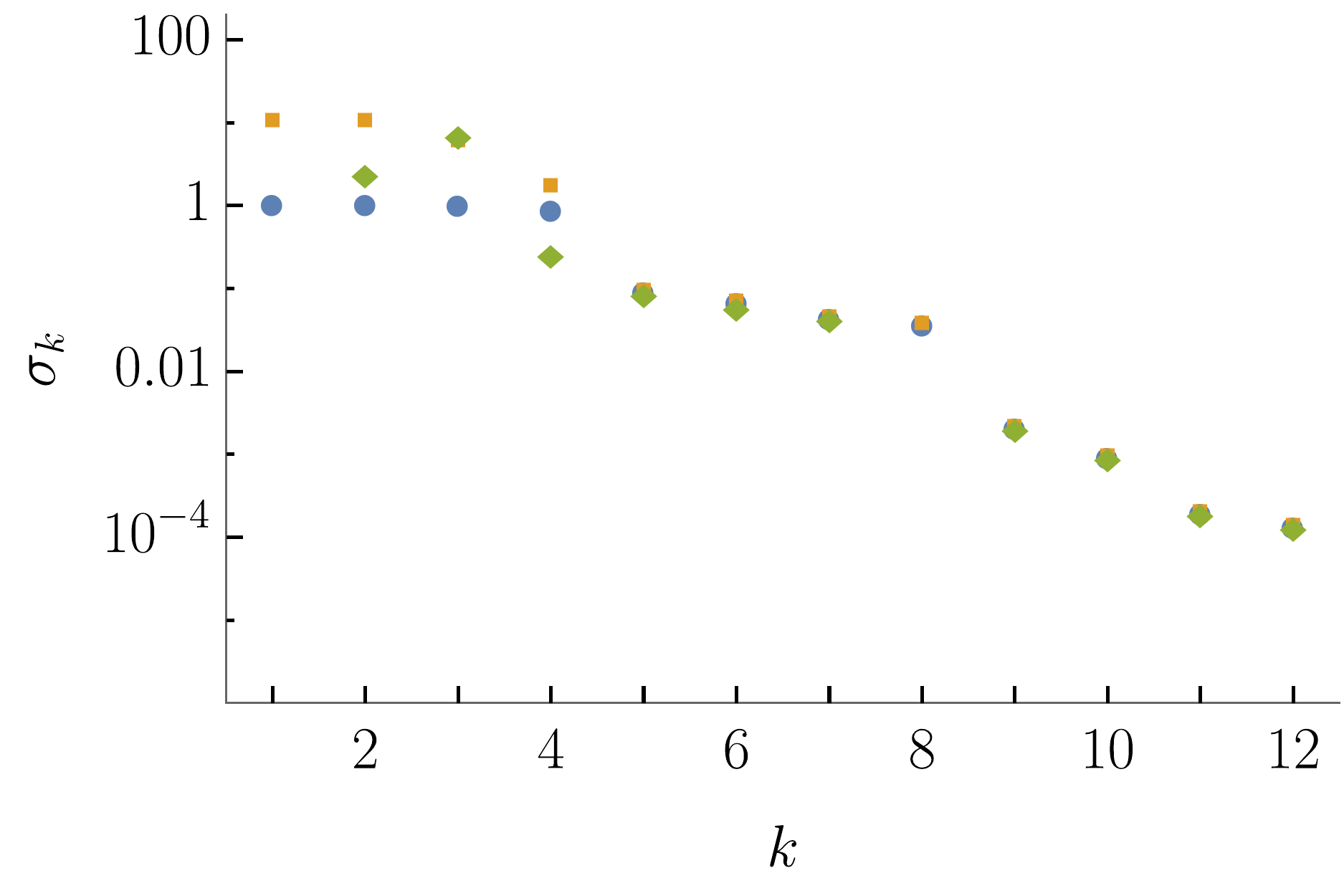}
 \caption{One sigma limits $\sigma_k$ on the $k = 1\ldots12$ eigenvectors $W_k$ of the covariance matrix for the least squares estimators.
 The blue circles and orange squares are the results of 12 parameter, regularized fits with $\kappa = 1$ and $10^{-2}$, respectively.
The green diamonds correspond to an unregularized, 10 parameter least squares fit.}
   \label{fig:eigenvalues}
\end{figure}

Results for $\sigma_k$ are reported in Fig.~\ref{fig:eigenvalues}.
The blue circles and orange squares are the results of 12 parameter, regularized fits with $\kappa = 1$ and $10^{-2}$, respectively.
From these fits we see that eigenvectors 1 and 2 are blind directions in parameter space as far as Higgs boson signal strengths are concerned.
A direction $k$ is called blind if $\sigma_k = 1 / \sqrt{\kappa}$, independent of the choice of $\kappa$.
An advantage of this approach is that it can quickly pick out these blind directions.
Explicit expression for the eigenvectors in the $\kappa = 1$ case are given in Appendix~\ref{sec:evec}.
Additional plots are presented in Appendix~\ref{sec:plots}.

In comparison, the green diamonds correspond to an unregularized, 10 parameter least squares fit.
The two parameters removed from Eq.~\eqref{eq:params} are $c_T$ and $g_2 c_W + g_1 c_B$, which are two linear combinations of parameters appearing in eigenvectors 1 and 2 of the regularized fits.
This is not a unique choice, but removing these combinations of parameters forces the oblique parameters $S$ and $T$~\cite{Peskin:1991sw} to be zero at tree level.
However we caution that when there are many operators that can potentially be non-zero, $c_T$ and $g_2 c_W + g_1 c_B$ need not be tiny to be consistent with EWPD~\cite{Han:2004az, Grinstein:2013vsa, Berthier:2015gja, SanchezColon:1998xg, Kilian:2003xt, Grojean:2006nn}.

When regularization is not important, eigenvectors 5 through 12, there is excellent agreement between the different cases.
The bounds on eigenvectors 9 through 12 are at the level of a few permille or stronger.
These eigenvectors are composed almost exclusively of $c_b$, $c_{\tau}$, $c_g$, and $c_{\gamma}$, respectively.
Weaker, percent level bounds, are found for eigenvectors 5 through 8.
The third eigenvector is almost entirely composed of $c_6$.
Not surprisingly the associated bound is weak, especially in the non- and weakly-regularized cases, while the regularization with $\kappa = 1$ makes this bound artificially stronger.
The correlation between the eigenvectors in regularized and unregularized cases is not exact, so a comparison of eigenvalues 2 and 4 is approximate.

The purpose of the regulation parameter is to the parameters of interest from becoming too large. 
Specifically, the standard deviations of the least squares estimators are regulated to have a maximum size
\begin{equation}
\label{eq:dc}
\Delta c_i \lsim \frac{1}{\sqrt{\kappa}} .
\end{equation}
Based on this, and given the normalization of the operators in Eq.~\eqref{eq:Lgen}, a choice of $\kappa$ can be seen as imposing a prior assumption of the lowest possible scale of BSM physics, $\Lambda_{\text{min}}$, or as imposing an upper limit on a signal strength from an experimental measurement.
For example, if $\kappa$ is taken to be 1 (10) then from~\eqref{eq:Lgen},~\eqref{eq:dc} we have $\Lambda_{\text{min}} \sim v\, (\Lambda_{\text{min}} \sim 800$~GeV).
The choice of $\kappa = 1$ can be seen as minimally enforcing the convergence of the EFT.
However this interpretation depends on the normalization of the operators in Eq.~\eqref{eq:Lgen}, with a different normalization generally leading to a different interpretation.
A choice of the regularization parameter less than one could be used to enforce an experimental upper limit on a process that is not yet well measured, such as double Higgs boson production~\cite{ATLAS-CONF-2016-049, CMS-PAS-HIG-17-008} or Higgs boson decay to a $Z$ boson and a photon~\cite{Chatrchyan:2013vaa, Aad:2015gba, Aaboud:2017uhw}.
However we do not pursue this approach, opting to include $h \to Z \gamma$ signal strengths in our fits, and making predictions for $g g \to h h$.
The interpretation also depends on the structure of the regularization matrix, $\kappa_{ij}$.
For instance, if one assumed a UV theory that is strongly-coupled, it might make more sense to relate the entries of $\kappa_{ij}$ to the size of the coefficients expected from na\"{i}ve dimensional analysis~\cite{Manohar:1983md}, rather than taking $\kappa_{ij}$ to be proportional to the identity matrix.

\subsection{Model-Independence of the Eigensystem}
Another way to understand the results is to look at two-dimensional profiles of the fits.
We focus on $c_{\gamma}$ and $c_g$ in what follows.
In doing so we demonstrate that the eigensystem has a certain amount of model-independence to it.
It would be interesting to investigate exactly how model-independent the eigensystem is.
The key point discussed below is that marginalized allowed regions of parameter space in the parameter basis are sensitive to assumptions about the UV physics, whereas in the eigenbasis this is not the case.

\begin{figure}
  \centering
\subfloat{\includegraphics[width=0.49\textwidth]{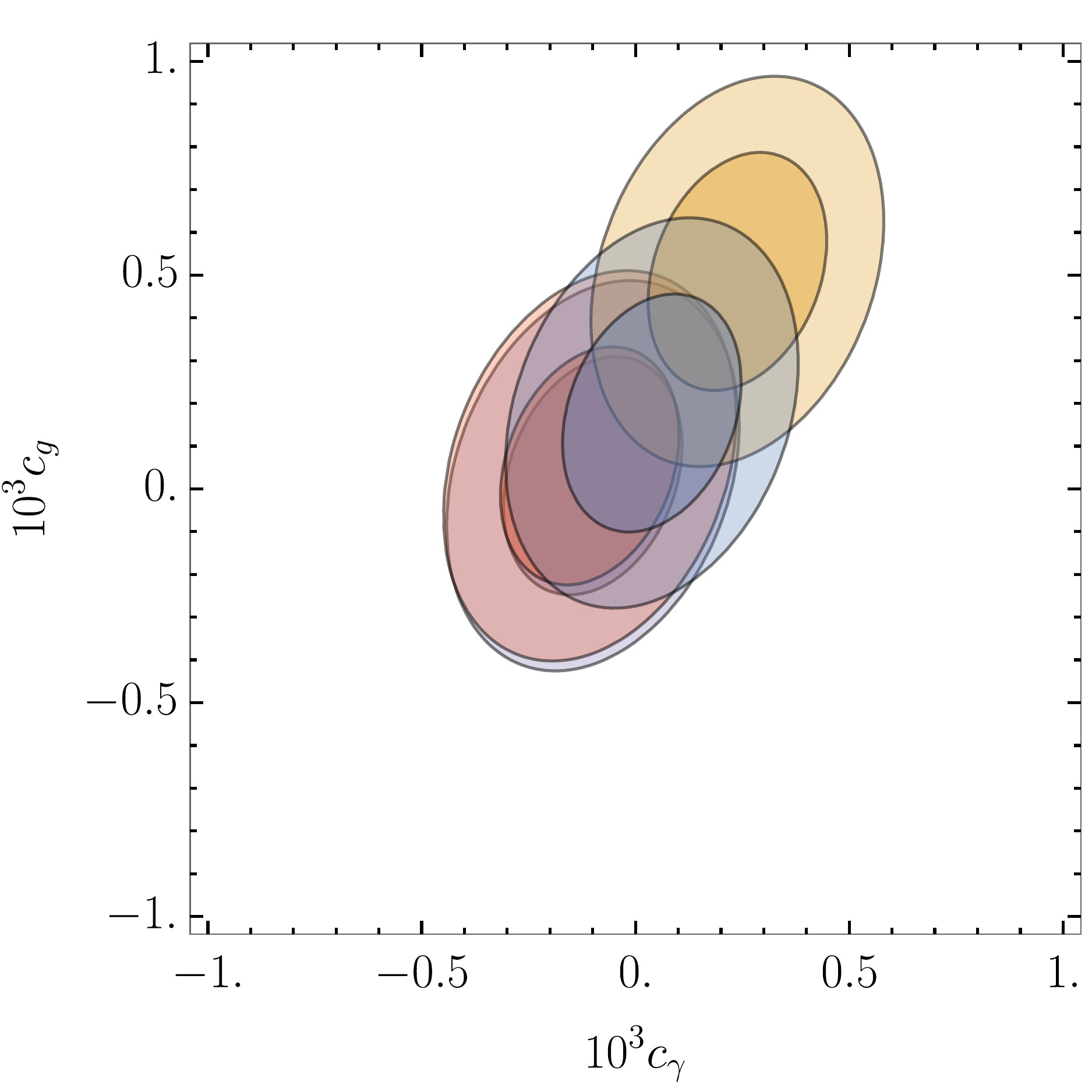}}\,
\subfloat{\includegraphics[width=0.49\textwidth]{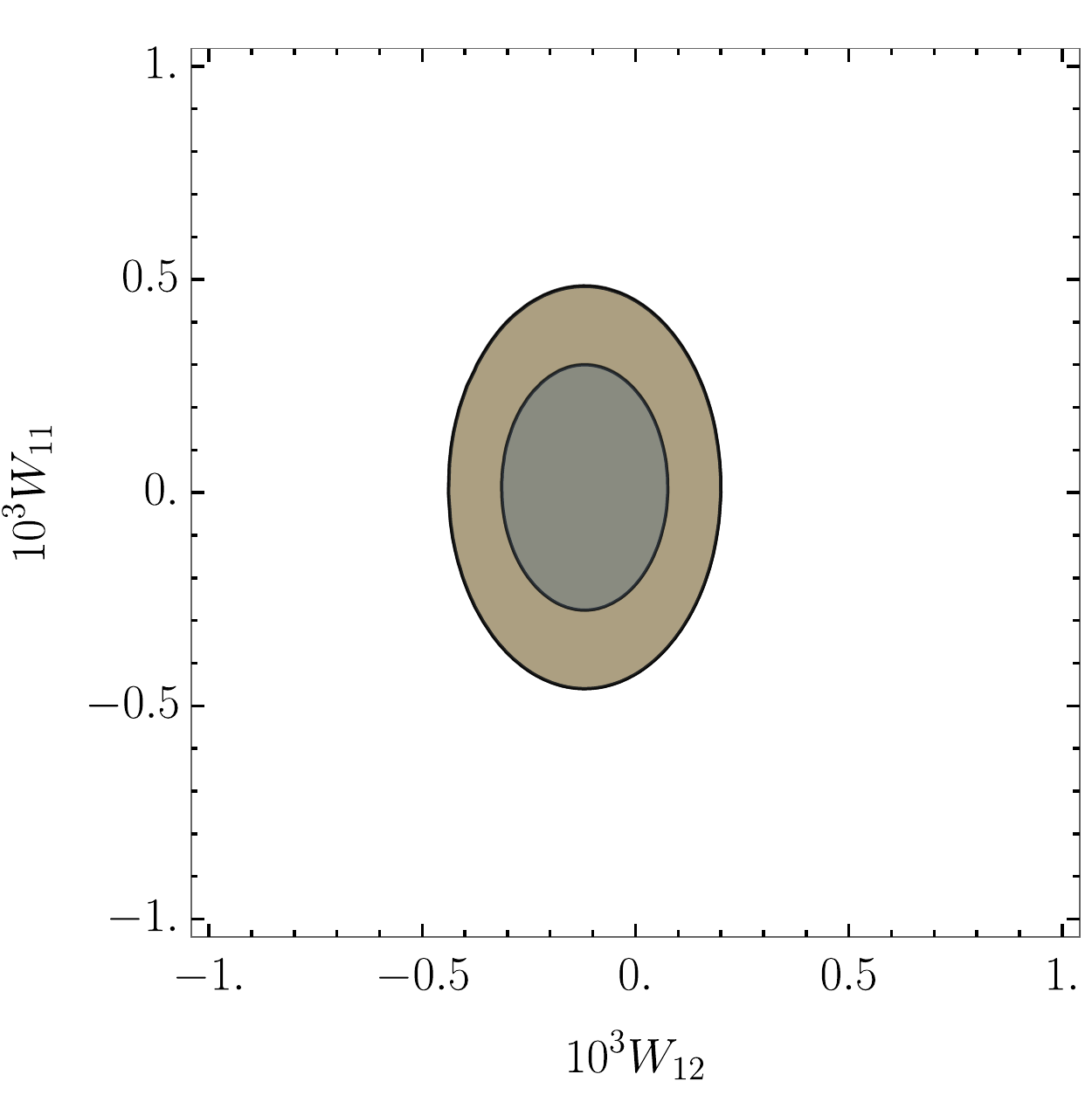}}
 \caption{(Left:) Preferred parameter space in the $c_g - c_{\gamma}$ plane based on the criteria $\Delta \chi^2 =$ 2.30 (darker shading) and 6.18 (lighter shading). 
The central values of $c_{\gamma}$ and $c_g$ depend on the assumption of what additional parameters may be non-zero and how large they can be, while the variances of and correlation between $c_{\gamma}$ and $c_g$ are the same in all cases.
See the text for details about the five scenarios. 
(Right:) In contrast, when the same fits are presented in terms of the eigenvectors $W_{11}$ and $W_{12}$, perfect agreement between the five cases is found. 
This indicates the eigensystem has a degree of model-independence to it.}
  \label{fig:ellipses}
\end{figure}

The left panel of Fig.~\ref{fig:ellipses} shows the one and two sigma preferred values for $c_{\gamma}$ and $c_g$ -- these regions are defined by $\Delta \chi^2 = 2.30$ and 6.18, respectively, with all other parameters fixed to their central values -- for five scenarios to be defined.
The blue and orange regions correspond to regularized fits with all 12 parameters, and $\kappa = 1$ and $10^{-2}$, respectively.
Darker (lighter) shading indicates the one (two) sigma allowed region.
The red region is an unregularized least squares fit where the only two non-zero parameters are $c_{\gamma}$ and $c_g$.
Furthermore the purple region is also an unregularized least squares fit, but where the four parameters are non-zero $\{c_{\gamma},\, c_g,\, c_{HW},\, c_{HB}\}$, and where $c_{HW}$, $c_{HB}$ are marginalized over.
There is a noticeable lack of agreement between these different scenarios as to what are the preferred central values of $c_{\gamma}$ and $c_g$ are.
In fact, for the fifth scenario, the unregularized 10 parameter fit described above, the preferred central values do not show up in the range of parameters plotted in Fig.~\ref{fig:ellipses}.
However note that the variances of and correlation between $c_{\gamma}$ and $c_g$ are the same in all five cases.

To understand what is happening here consider the right panel of Fig.~\ref{fig:ellipses}.
This shows the same five fits, but in the plane of the eigenvectors $W_{11}$ and $W_{12}$, rather than the parameters $c_{g}$, $c_{\gamma}$.
All five fits agree perfectly as to what the preferred region is in this case.
That such a difference occurs is interesting because $W_{12}$ and $W_{11}$ are composed almost exclusively $c_{\gamma}$ and $c_g$, respectively; $w_{12,\gamma} \approx w_{11,g} \approx 0.93$.
The difference, or lack thereof, between the central values in a given scenario occurs because when the additional parameters are fixed to their central values, which in turn forces $c_{\gamma}\, (c_g)$ away from the central value of $W_{12}\, (W_{11})$.
In the case of the fit with only $c_{\gamma}$ and $c_g$ non-zero from the start, there are no additional parameters, and allowed contours in the $c_g - c_{\gamma}$ and $W_{11} - W_{12}$ plane are identical up to the rotating induced in going from one basis to the other.
This shows that eigensystem is a fairly model-independent quantity.
It depends only on the SMEFT framework, and in the particular cases of $W_{11}$ or $W_{12}$, for example, that the parameters $c_{\gamma}$ or $c_g$ can be non-zero, but with no additional assumption about which parameters may or may not be non-zero.

\subsection{Predictions}
An advantage of having estimates for all of the coefficients under consideration is that predictions can be made for observables that have not been measured yet.
For example, a prediction can be made for the total width of the Higgs boson.
We find, using only Run-1 results,
\begin{equation}
\frac{\Gamma_{SMEFT, h}}{\Gamma_{SM, h}} \simeq 0.5 \pm 0.4 \quad \text{(Run-1).}
\end{equation}
The Higgs decay rate to bottom quarks -- the largest branching fraction in the SM -- was measured to be low during Run-1 of the LHC, which explains why this value is below the SM prediction. 
Adding results from Run-2, which are closer to the SM prediction, we instead find
\begin{equation}
\frac{\Gamma_{SMEFT, h}}{\Gamma_{SM, h}} \simeq 0.9 \pm 0.3 \quad \text{(Run-1+Run-2).}
\end{equation}
As expected the central value is higher, and now the prediction for the width of the Higgs boson in the SMEFT is consistent with the SM prediction.
In addition, predictions can also be made for double Higgs boson production.
The CMS Run-2 upper limit for double Higgs production is 19.2 times the SM prediction, at the 95\% CL~\cite{CMS-PAS-HIG-17-008}.
The upper limit we derive for double Higgs production in the SMEFT in the most general case is not competitive with the experimental upper limit, indicating that the experiments are currently sensitive to non-resonant double Higgs production.
Explicit bounds on double Higgs production in the SMEFT in the general case are shown in Appendix~\ref{sec:plots}, Fig.~\ref{fig:doublehiggs} specifically.
On the other hand, in specific scenarios tight bounds on double Higgs production can be derived.
For example, setting $c_6$ to zero we find $\sigma_{SMEFT}(g g \to h h) / \sigma_{SM}(g g \to h h) \simeq 1.4 \pm 0.4$.

\subsection{Electroweak Baryogenesis}
\begin{figure}
  \centering
\includegraphics[width=0.65\textwidth]{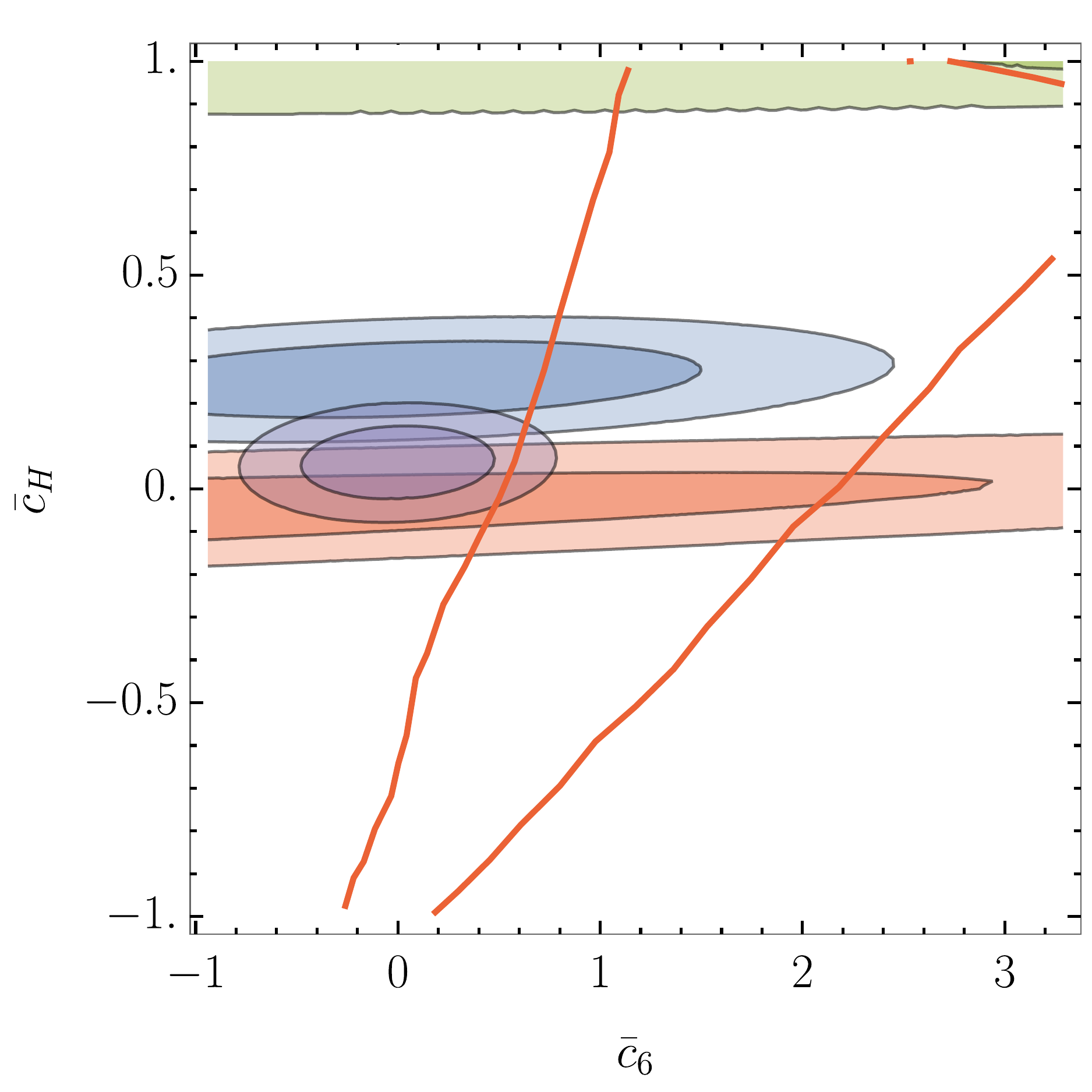}
 \caption{Constraints on the parameter space relevant for EW baryogenesis in the SMEFT.
 A first order phase transition occurs in the wedge bounded by the red lines~\cite{Damgaard:2015con}, and viable parameter space still exists.
 The purple ellipses are the result of a regularized fit with $\kappa = 10$, which is not consistent with a first order phase transition at the one sigma level.
A regularization parameter of 10 approximately corresponds to an effective scale of 800~GeV, reinforcing the notion that successful EW baryogenesis requires fairly low scale BSM physics~\cite{Grojean:2004xa, Kobakhidze:2015xlz}.}
  \label{fig:ewbaryogenesis}
\end{figure}

The trilinear Higgs coupling plays an important role in not only double Higgs production, but also in EW baryogenesis.
To investigate the constraints on EW baryogenesis in the SMEFT we switch to a more common notation:
\begin{equation}
c_H = \frac{1}{2} \bar{c}_H , \quad c_6 = - \frac{m_h^2}{2 v^2} \bar{c}_6 .
\end{equation}
Assuming temperature dependence only in the Higgs mass parameter, requiring a first order phase transition yields the analytic bound~\cite{Grojean:2004xa, Huang:2015izx, Huang:2015tdv, Jain:2017sqm}
\begin{equation}
\frac{2}{3} < \bar{c}_6 < 2 .
\end{equation}
Ref.~\cite{Damgaard:2015con} went further and considered the viable parameter space for a first order phase transition in the $\bar{c}_H - \bar{c}_6$ plane.
This parameter space is bounded by our fits, and is shown in Fig.~\ref{fig:ewbaryogenesis}.
A first order phase transition occurs in the wedge bounded by the red lines~\cite{Damgaard:2015con}.
The blue and green ellipses give the favored parameter space from the 12 parameter regularized fit with $\kappa = 1$, and the 10 parameter unregularized fit, respectively.
Darker and lighter shading again correspond to $\Delta \chi^2 = 2.30$ and 6.18, respectively.
The parameter space shaded red is the favored result of standard two parameter fit assuming only $\bar{c}_H$ and $\bar{c}_6$ are non-zero.
Although the parameter space is constrained, EW baryogenesis in the SMEFT is still viable provided the cutoff scale is not too large.
This is explicitly demonstrated by a regularized fit with $\kappa = 10$, shown by the purple ellipses, which is not consistent with a first order phase transition at the one sigma level.
A regularization parameter of 10 approximately corresponds to an effective scale of 800~GeV, reinforcing the notion that successful EW baryogenesis requires fairly low scale BSM physics~\cite{Grojean:2004xa, Kobakhidze:2015xlz}.

\subsection{Future Measurements}
One may wonder which experimental measurements would improve the global constraints the most.
To this end, we add to our fit one hypothetical future signal strength of $1.0 \pm 0.1$ for various Higgs boson observables, and see how this changes the fit.
This is quantified using the global determinant parameter of Ref.~\cite{Durieux:2017rsg}.
The GDP is defined in our notation as
\begin{equation}
\text{GDP} = \left(\prod_{j \subseteq k} \sigma_j^2\right)^{\tfrac{1}{m}} ,
\end{equation}
where $m$ is the total number of eigenvalues considered, which need not be all 12 in general.
In particular, we report the ratio of the GDP with this additional hypothetical measurement to the the GDP of the 55 measurement fit described previously.
We confirm that ratios of GDPs do not depend on the normalization of the operators.
The list of all observables that improve the constraints by themselves are given in Tab.~\ref{tab:GDP}.
We use the unregularized, 10 parameter fit in computing these GDPs.
Some of the observables, such as double Higgs production, are obvious candidates, but others are less well known to be important for future Higgs coupling constraints.
\begin{table}
\centering
 \begin{tabular}{| c | c || c | c |}
 \hline 
Observable & GDP ratio & Observable & GDP ratio  \\ \hline 
$g g \to h h$ & 0.37 & $W h$, $h \to Z Z^*$ & 0.96 \\ \hline 
$h \to Z \gamma$ & 0.71 & VBF, $h \to b \bar{b}$ & 0.98 \\ \hline  
$h \to c \bar{c}$ & 0.80 & $\Gamma_h$ & 0.98 \\ \hline 
$h \to \mu^+ \mu^-$ & 0.80 & $Z h$, $h \to \tau^+ \tau^-$ & 0.99 \\ \hline  
$t t h$, $h \to Z Z^*$ & 0.93 & $t t h$, $h \to b \bar{b}$ & 0.99 \\ \hline  
$Z h$, $h \to Z Z^*$ & 0.94 & $g g$F, $h \to b \bar{b}$ & 0.99 \\ \hline   
 \end{tabular}
  \caption{Improvement in the global constraints by adding one hypothetical signal strength of $1.0 \pm 0.1$ to the fit for various Higgs observables.
  The improvement is quantified using the ratio of the GDPs of the fits with/without the hypothetical measurement.}
  \label{tab:GDP}
\end{table}

\section{Summary}
\label{sec:sum}
In this work we performed an SMEFT parameter fit with an emphasis on a statistical technique aimed at tackling the issue of the large number of parameters.
The technique we used is a regularized linear regression, where a positive definite function of the parameters of interest is added to the usual cost function.
This prevents the fit from falling into an overfit solution, and, in principle, allows information to be obtained about any number of parameters.
A cross-validation was performed to try to determine the optimal value of the regularization parameter to use.
The cross-validation instead selected the SM as the best fit, so we contented ourselves to performing regularized fits with multiple choices for the regularization parameter, and examined how much regulator dependence various quantities had. 
As proof of principle of this technique we applied it to fitting Higgs boson signal strengths, including the latest Run-2 results.
We emphasized presenting results in terms of the eigensystem of the covariance matrix of the least squares estimators as it has a degree model-independent to it.
We showed the SMEFT predicts the total width of the Higgs boson, which is not yet directly measured, to be consistent with the SM prediction, and that the ATLAS and CMS experiments at the LHC are currently sensitive to non-resonant double Higgs boson production.
We derived constraints on the viable parameter space for EW baryogenesis in the SMEFT, and reinforce the notion that a first order phase transition requires fairly low scale BSM physics.
We studied which future experimental measurements would improve the global constraints on the Higgs sector of the SMEFT the most.
This is quantified using ratios of the GDP, which has a natural interpretation in terms of the eigensystem of the covariance matrix.
We expect this technique to be of use to practitioners of both bottom-up and top-down approaches to EFTs.

\begin{acknowledgments}
We are grateful to Marco Battaglia, Olaf Behnke, Roberto Contino, Laura Covi, Sally Dawson, Prerit Jaiswal, Alexander Kusenko, Tania Robens, David Stone, Michael Trott, Susanne Westhoff, and Tevong You for useful discussions.
We thank the Galileo Galilei Institute for Theoretical Physics, Sapienza Universit\`{a} di Roma, and Scuola Normale Superiore for their hospitality during the completion of this work; and King's College and the University of Bristol for opportunities to present preliminary versions of this work. 
This work was supported by the United States Department of Energy under Grant Contract DE-SC0012704.
\end{acknowledgments}

\appendix
\section{Experimental Results}
\label{sec:exp}
The experimental results used in this analysis from Run-1 of the LHC are given in Table~\ref{tab:exp1}.
Similarly, the ATLAS and CMS Run-2 results can be found in Tables~\ref{tab:exp2a} and~\ref{tab:exp2c}, respectively.

\begin{table}
\centering
 \begin{tabular}{| c | c | c || c | c | c |}
 \hline 
Production & Decay & Signal Strength & Production & Decay & Signal Strength  \\ \hline 
$gg$F & $\gamma\gamma$ & $1.10^{+0.23}_{-0.22}$ & $Wh$ & $bb$ & $1.0\pm0.5$ \\ \hline
$gg$F & $ZZ$ & $1.13^{+0.34}_{-0.31}$ & $Zh$ & $\gamma\gamma$ & $0.5^{+3.0}_{-2.5}$ \\ \hline
$gg$F & $WW$ & $0.84\pm0.17$ & $Zh$ & $WW$ & $5.9^{+2.6}_{-2.2}$ \\ \hline
$gg$F & $\tau\tau$ & $1.0\pm0.6$ & $Zh$ & $\tau\tau$ & $2.2^{+2.2}_{-1.8}$ \\ \hline
VBF & $\gamma\gamma$ & $1.3\pm0.5$ & $Zh$ & $bb$ & $0.4\pm0.4$ \\ \hline
VBF & $ZZ$ & $0.1^{+1.1}_{-0.6}$ & $tth$ & $\gamma\gamma$ & $2.2^{+1.6}_{-1.3}$ \\ \hline
VBF & $WW$ & $1.2\pm0.4$ & $tth$ & $WW$ & $5.0^{+1.8}_{-1.7}$ \\ \hline
VBF & $\tau\tau$ & $1.3\pm0.4$ & $tth$ & $\tau\tau$ & $-1.9^{+3.7}_{-3.3}$ \\ \hline
$Wh$ & $\gamma\gamma$ & $0.5^{+1.3}_{-1.2}$ & $tth$ & $bb$ & $1.1\pm1.0$ \\ \hline
$Wh$ & $WW$ & $1.6^{+1.2}_{-1.0}$ & $pp$ & $\mu\mu$ & $0.1\pm2.5$ \\ \hline
$Wh$ & $\tau\tau$ & $-1.4\pm1.4$ & $pp$ & $Z\gamma$ & $2.7^{+4.6}_{-4.5}$ \\ \hline
 \end{tabular}
  \caption{Run-1 experimental results used in this work. 
  The $Z\gamma$ result is from ATLAS~\cite{Aad:2015gba}.
  CMS does not provide a signal strength for $h \to Z \gamma$ although their 95\% CL upper limit is stronger~\cite{Chatrchyan:2013vaa} than the ATLAS Run-1 result. 
  All other results are taken from the combined ATLAS+CMS analysis of Ref.~\cite{Khachatryan:2016vau} with correlations taken into account.}
  \label{tab:exp1}
\end{table}
\begin{table}
\centering
 \begin{tabular}{| c | c | c | c || c | c | c | c |}
 \hline 
Production & Decay & Signal Strength & Reference & Production & Decay & Signal Strength  & Reference \\ \hline 
$pp$ & $\mu\mu$ & $-0.1\pm1.4$ & \cite{Aaboud:2017ojs} & $gg$F & $ZZ$ & $1.11^{+0.25}_{-0.22}$ & \cite{ATLAS-CONF-2017-043}  \\ \hline
$Wh$ & $bb$ & $1.35^{+0.68}_{-0.59}$ & \cite{Aaboud:2017xsd} & VBF & $ZZ$ & $4.0^{+1.8}_{-1.5}$ & \cite{ATLAS-CONF-2017-043}  \\ \hline
$Zh$ & $bb$ & $1.12^{+0.50}_{-0.45}$ & \cite{Aaboud:2017xsd} & VBF & $WW$ & $1.7^{+1.2}_{-0.9}$ & \cite{ATLAS-CONF-2016-112}  \\ \hline
$gg$F & $\gamma\gamma$ & $0.80^{+0.19}_{-0.18}$ & \cite{ATLAS-CONF-2017-045} & $Wh$ & $WW$ & $3.2^{+4.4}_{-4.2}$ & \cite{ATLAS-CONF-2016-112}  \\ \hline
VBF & $\gamma\gamma$ & $2.1\pm0.6$ & \cite{ATLAS-CONF-2017-045} & $tth$ & $2\ell\, 0\tau_h$ & $4.0^{+2.1}_{-1.7}$ & \cite{ATLAS-CONF-2016-058}  \\ \hline
$Vh$ & $\gamma\gamma$ & $0.7^{+0.9}_{-0.8}$ & \cite{ATLAS-CONF-2017-045} & $tth$ & $2\ell\, 1\tau_h$ & $6.2^{+3.6}_{-2.7}$ & \cite{ATLAS-CONF-2016-058}  \\ \hline
$tth$ & $\gamma\gamma$ & $0.5\pm0.6$ & \cite{ATLAS-CONF-2017-045} & $tth$ & $3\ell$ & $0.5^{+1.7}_{-1.6}$ & \cite{ATLAS-CONF-2016-058}  \\ \hline
$pp$ & $Z\gamma$ & $1.3\pm2.6$ & \cite{Aaboud:2017uhw} & & & &   \\ \hline
 \end{tabular}
  \caption{Run-2 ATLAS results used in this work.
  We estimate the signal strength for $h \to Z \gamma$ from Ref.~\cite{Aaboud:2017uhw}, which states the upper limit for this process is 6.6 times the SM rate at 95\% CL and that the significance of the measurement is $0.5 \sigma$.}
  \label{tab:exp2a}
\end{table}
\begin{table}
\centering
 \begin{tabular}{| c | c | c | c || c | c | c | c |}
 \hline 
Production & Decay & Signal Strength & Reference & Production & Decay & Signal Strength  & Reference \\ \hline 
$gg$F & $ZZ$ & $1.20\pm0.20$ & \cite{Sirunyan:2017exp} & $gg$F & $\gamma\gamma$ & $1.11^{+0.19}_{-0.18}$ & \cite{CMS-PAS-HIG-16-040} \\ \hline
0-jet & $\tau\tau$ & $0.84\pm0.89$ & \cite{Sirunyan:2017khh} & VBF & $\gamma\gamma$ & $0.5^{+0.6}_{-0.5}$ & \cite{CMS-PAS-HIG-16-040} \\ \hline
VBF & $\tau\tau$ & $1.11^{+0.34}_{-0.35}$ & \cite{Sirunyan:2017khh} & $Vh$ & $\gamma\gamma$ & $2.3^{+1.1}_{-1.0}$ & \cite{CMS-PAS-HIG-16-040} \\ \hline
$tth$ & $2\ell$ & $1.7^{+0.6}_{-0.5}$ & \cite{CMS-PAS-HIG-17-004} & $tth$ & $\gamma\gamma$ & $2.2^{+0.9}_{-0.8}$ & \cite{CMS-PAS-HIG-16-040} \\ \hline
$tth$ & $3\ell$ & $1.0^{+0.8}_{-0.7}$ & \cite{CMS-PAS-HIG-17-004} & 0-jet & $WW$ & $0.9^{+0.4}_{-0.3}$ & \cite{CMS-PAS-HIG-16-021} \\ \hline
$tth$ & $4\ell$ & $0.9^{+2.3}_{-1.6}$ & \cite{CMS-PAS-HIG-17-004} & VBF & $WW$ & $1.4\pm0.8$ & \cite{CMS-PAS-HIG-16-021} \\ \hline
$tth$ & $\tau\tau$ & $0.72^{+0.62}_{-0.53}$ & \cite{CMS-PAS-HIG-17-003} & $Wh$ & $WW$ & $-1.4\pm1.5$ & \cite{CMS-PAS-HIG-16-021} \\ \hline
$Wh$ & $bb$ & $1.7\pm0.7$ & \cite{CMS-PAS-HIG-16-044} & $Vh$ & $WW$ & $2.1^{+2.3}_{-2.2}$ & \cite{CMS-PAS-HIG-16-021} \\ \hline
$Zh$ & $bb$ & $0.9\pm0.5$ & \cite{CMS-PAS-HIG-16-044} & $tt$ & $bb$ & $-0.19^{+0.82}_{-0.81}$ & \cite{CMS-PAS-HIG-16-038} \\ \hline
 \end{tabular}
  \caption{Run-2 CMS results used in this work.}
  \label{tab:exp2c}
\end{table}

\section{Eigenvectors}
\label{sec:evec}
The eigenvectors for the regularized fit with $\kappa = 1$ are
\begin{align}
W_1 &\simeq 0.99 c_B + 0.09 c_{HW} + 0.09 c_T - 0.08 c_W + 0.05 c_{HB} , \\
W_2 &\simeq 0.67 c_{HW} - 0.56 c_W - 0.36 c_B + 0.33 c_{HB} - 0.02 c_T , \nn \\
W_3 &\simeq 0.99 c_6 - 0.13 c_t + 0.05 c_W + 0.03 c_{HW} + 0.02 c_{HB} - 0.01 c_H , \nn \\
W_4 &\simeq 0.67 c_W + 0.45 c_{HW} + 0.38 c_H - 0.38 c_t + 0.24 c_{HB} - 0.09 c_6 , \nn \\
W_5 &\simeq 0.76 c_t + 0.40 c_W - 0.32 c_H + 0.24 c_{HW} + 0.22 c_T + 0.20 c_{HB} + 0.06 c_6 - 0.02 c_B , \nn \\
W_6 &\simeq 0.78 c_H + 0.51 c_t - 0.32 c_T - 0.10 c_W + 0.08 c_6 - 0.07 c_{HB}  - 0.05 c_{HW} + 0.03 c_b + 0.03 c_B , \nn \\
W_7 &\simeq 0.87 c_{HB} - 0.48 c_{HW} +0.09 c_H + 0.09 c_T - 0.06 c_W - 0.03 c_t + 0.01 c_b , \nn \\
W_8 &\simeq 0.91 c_T + 0.34 c_H - 0.16 c_{HB} - 0.11 c_W - 0.08 c_B - 0.03 c_{HW} + 0.03 c_b + 0.01 c_6 , \nn \\
W_9 &\simeq 0.97 c_b + 0.24 c_{\tau} - 0.07 c_g +0.04 c_{\gamma} - 0.03 c_H + 0.02 c_W + 0.01 c_{HW} - 0.01 c_t - 0.01 c_T , \nn \\
W_{10} &\simeq 0.97 c_{\tau} - 0.24 c_b + 0.05 c_g , \nn \\
W_{11} &\simeq 0.93 c_g + 0.35 c_{\gamma} + 0.06 c_b - 0.03 c_{\tau} , \nn \\
W_{12} &\simeq 0.93 c_{\gamma} - 0.35 c_g - 0.07 c_b . \nn
\end{align}
All 12 parameters contribute to each eigenvector, but only percent level or higher contributions are shown.
The blind directions are (two linear combinations of) eigenvectors 1 and 2.

\section{Additional Plots}
\label{sec:plots}
\begin{figure}
  \centering
\includegraphics[width=0.65\textwidth]{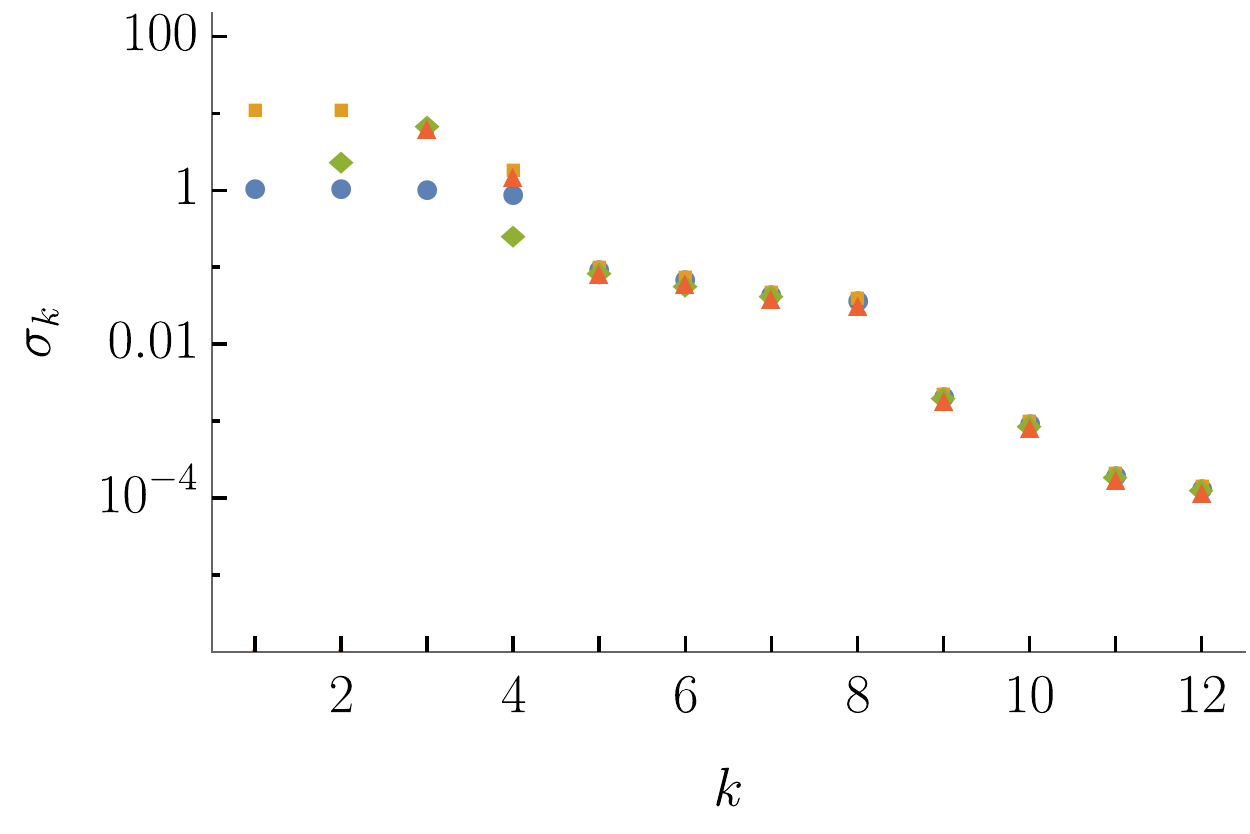}
 \caption{The same as Fig.~\ref{fig:eigenvalues}, but with the results of the pseudoinverse fit included as the red triangles.
 The eigenvalues of the blind directions are zero when using the pseudoinverse with regularization, and none of the other eigenvalues have regulator dependence.}
  \label{fig:evA}
\end{figure}

Another way to visualize the blind directions in the fit is to use the pseudoinverse without regularization to invert the Fisher information.
Given a matrix $A$, its pseudoinverse, $A^p$, is defined as
\begin{equation}
\label{eq:pseudo}
A A^p A = A,
\end{equation}
as opposed to $A^{-1} A = \mathbbm{1}$ for the case of the (genuine) inverse.
In this case the eigenvalues of the blind directions are zero, and there is no regulator dependence in any of the other eigenvalues.
This is shown in Fig.~\ref{fig:evA}.

\begin{figure}
  \centering
\includegraphics[width=0.65\textwidth]{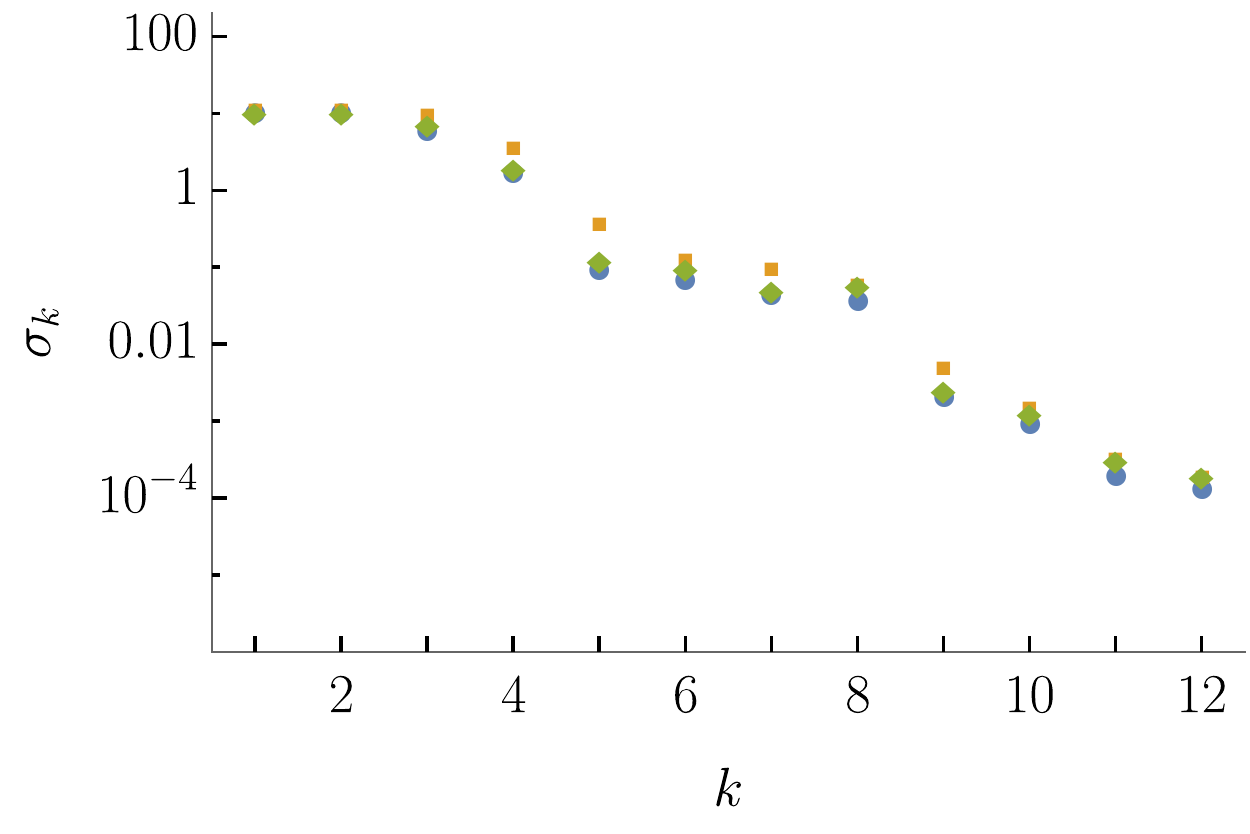}
 \caption{Results of regularized fits with $\kappa = 10^{-2}$ using only Run-1 results (orange squares), only Run-2 results (green diamonds), and both runs (blue circles).
 Many of the bounds are now driven by the Run-2 results, but the contributions from the Run-1 measurements are still important.}
  \label{fig:run1vsrun2}
\end{figure}

The combination of Run-1 and Run-2 works exactly as expected, adding more data tightens the resulting bounds, see Fig.~\ref{fig:run1vsrun2}.
Many of the bounds are now driven by the Run-2 results, but the contributions from the Run-1 measurements are still important.
Furthermore, Run-2 shows more sensitivity to $c_6$ as a result of its improved $tth$ measurements.

\begin{figure}
  \centering
\subfloat{\includegraphics[width=0.49\textwidth]{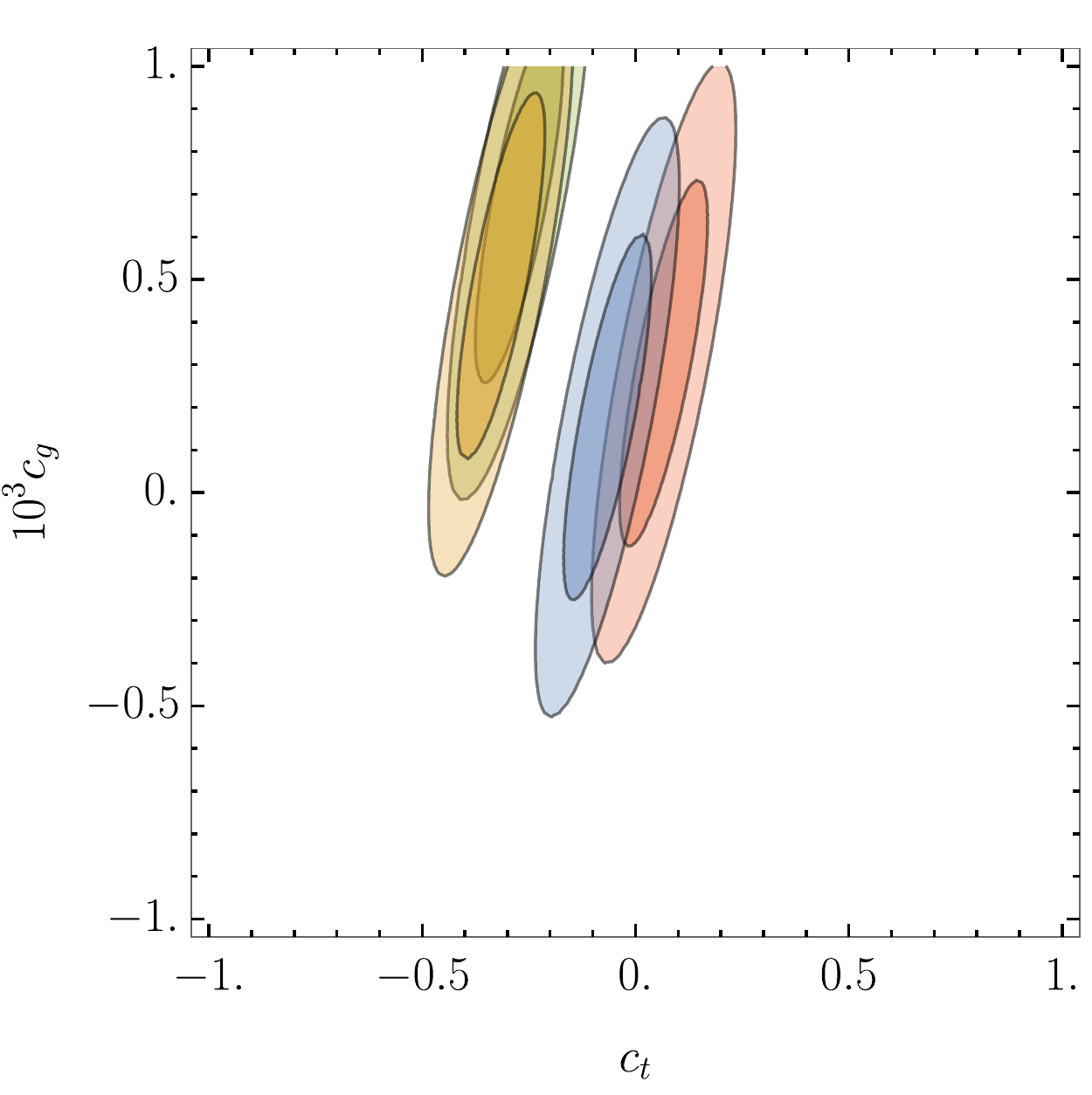}}\,
\subfloat{\includegraphics[width=0.49\textwidth]{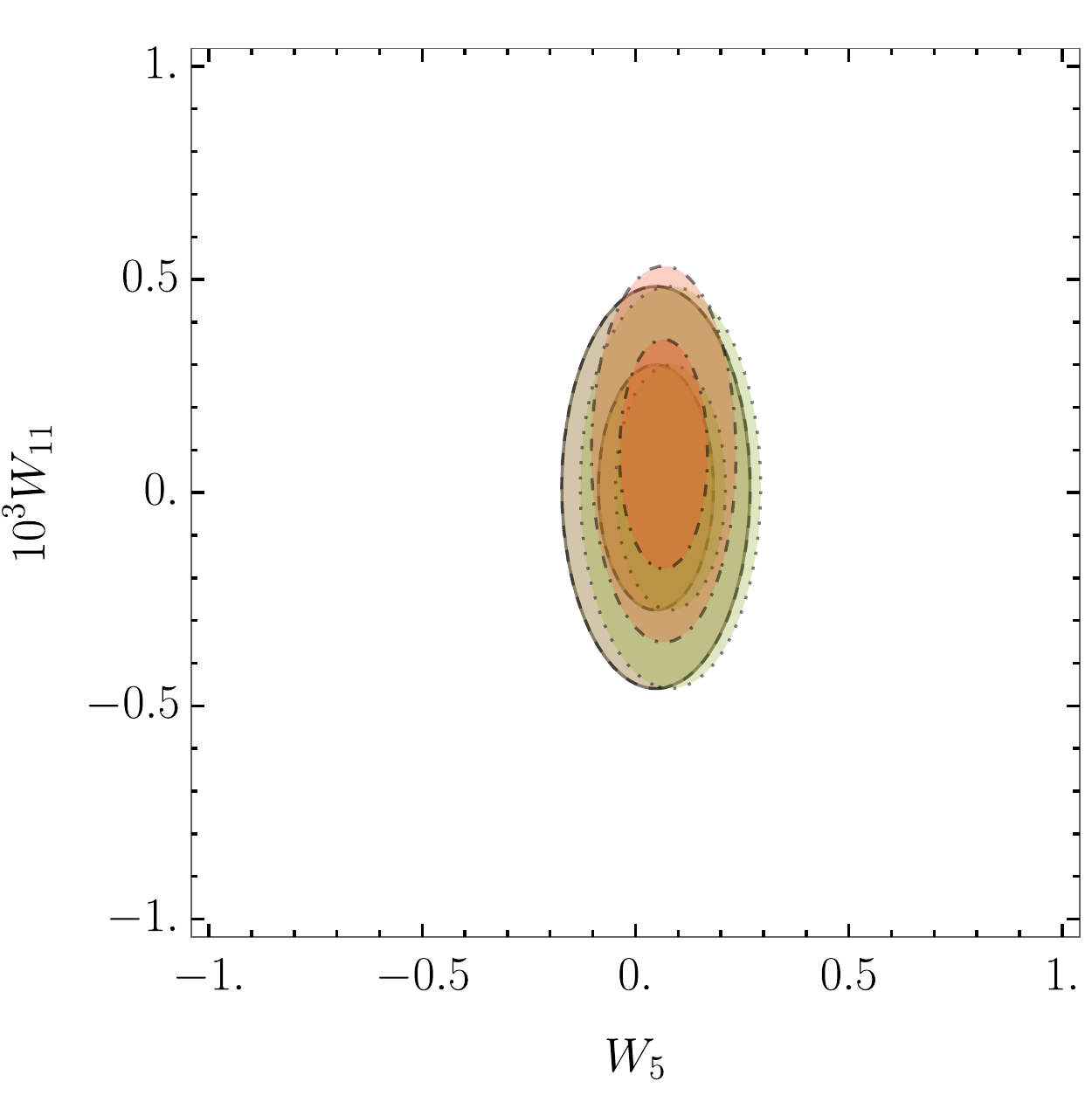}}
 \caption{The same as Fig.~\ref{fig:ellipses} but in the $c_g - c_t$ plane (left), and the $W_{11} - W_5$ plane (right).
 The correlation between the eigenvectors in the different scenarios is not perfect.
 Nevertheless the agreement between the different scenarios is better in the eigenvector basis.}
  \label{fig:cgct}
\end{figure}

There is better agreement between different scenarios, involving UV assumption or choice of regularization parameter, in the eigenvectors basis even when the correlation between the eigenvectors in the different scenarios is not perfect.
This is shown in Fig.~\ref{fig:cgct}, which is the same as Fig.~\ref{fig:ellipses} but in the $c_g - c_t$ plane (left), and the $W_{11} - W_5$ plane (right).

\begin{figure}
  \centering
\includegraphics[width=0.65\textwidth]{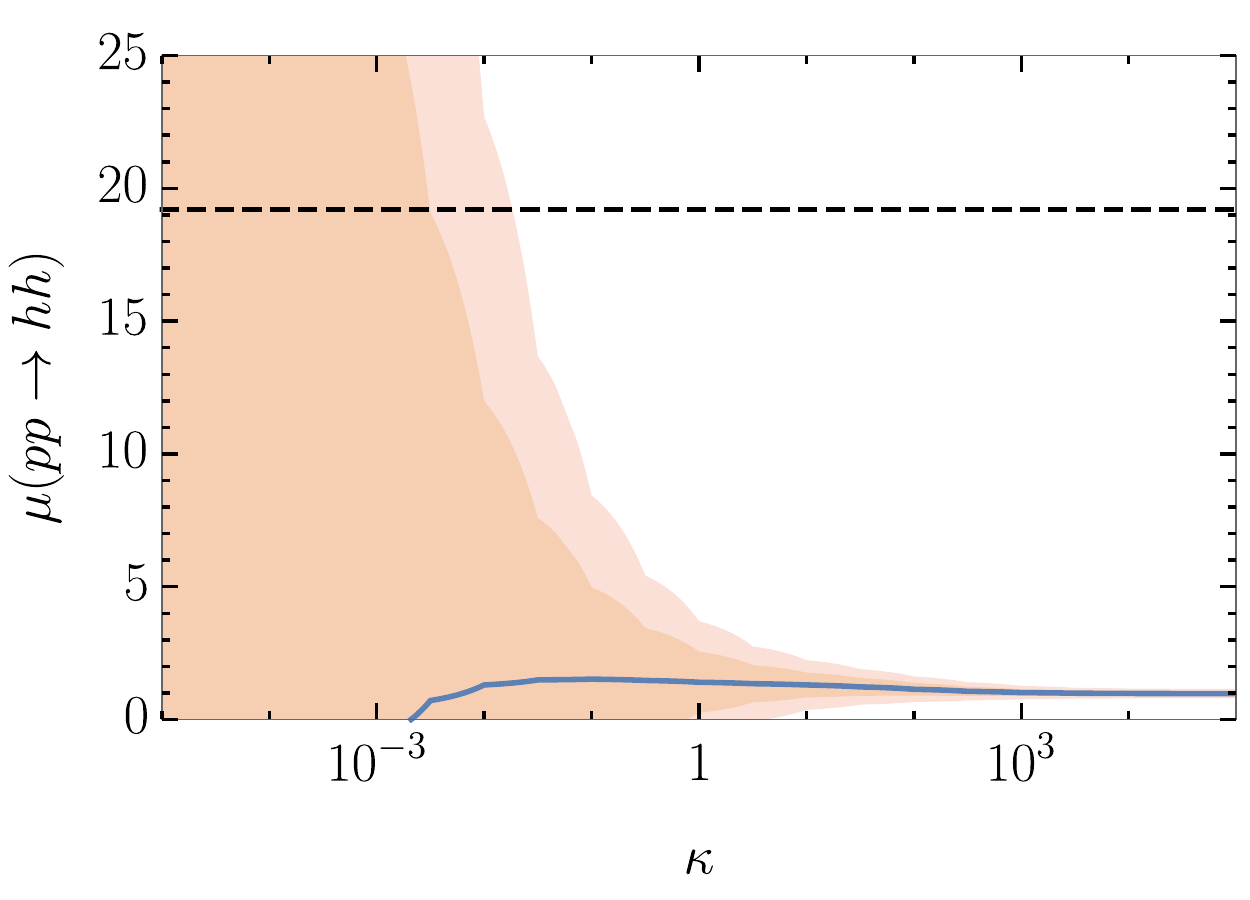}
 \caption{Bounds on double Higgs production, $\mu(p p \to h h) = \sigma(p p \to h h) / \sigma_{SM}(p p \to h h)$, when all 12 parameters are allowed to be non-zero as a function of the regularization parameter, $\kappa$. 
 The blue curve gives the best fit values, and the darker and lighter shaded regions are allowed at 1 and $2\sigma$, respectively.
 The dashed line is the experimental upper limit from CMS~\cite{CMS-PAS-HIG-17-008}.}
  \label{fig:doublehiggs}
\end{figure}

Bounds on double Higgs production, $\mu(p p \to h h) = \sigma(p p \to h h) / \sigma_{SM}(p p \to h h)$, when all 12 parameters are allowed to be non-zero as a function of the regularization parameter $\kappa$ are shown in Fig.~\ref{fig:doublehiggs}. 
 The blue line gives the best fit values, and the darker and lighter shaded regions are allowed at 1 and $2\sigma$, respectively.
 The dashed line is the experimental upper limit from CMS~\cite{CMS-PAS-HIG-17-008}.

\bibliographystyle{utphys}
\bibliography{SMEFTHiggsConstraints}

\end{document}